\documentclass[12pt]{article} 

\pdfoutput=1

\usepackage[a4paper,margin=25mm]{geometry}
\usepackage[normalem]{ulem}
\usepackage{color}

\usepackage[pdftex]{graphicx}
\usepackage{hyperref}

\usepackage[hang,small,bf]{caption}
\usepackage[subrefformat=parens]{subcaption}
\usepackage{amsmath,amssymb,bm}
\usepackage{physics}   
\usepackage{braket}
\usepackage{ascmac}
\usepackage{cite}
\usepackage{appendix}

\usepackage[compat=1.1.0]{tikz-feynhand}

\def\u1{\mathrm{U}(1)}

\newcommand{\ol}[1]{\overline{#1}}
\newcommand{\vp}{\varphi}
\newcommand{\mc}[1]{\mathcal{#1}}

\newcommand{\wt}[1]{\widetilde{#1}}

\graphicspath{{./Figures/}} 

\begin{document}
\setlength{\baselineskip}{0.7cm}
\begin{titlepage}
\begin{flushright}
NITEP 264
\end{flushright}
\vspace*{10mm}%
\begin{center}{\Large\bf
Pseudo-Nambu-Goldstone Dark Matter \\
\vspace*{2mm}
in Flux Compactification
}
\end{center}
\vspace*{10mm}
\begin{center}
{\large Kento Akamatsu}$^{a}$, 
{\large Takuya Hirose}$^{b}$, 
{\large Nobuhito Maru}$^{a,c}$ 
	and
{\large Akio Nago}$^{a}$
\end{center}
\vspace*{0.2cm}
\begin{center}
${}^{a}${\it
Department of Physics, Osaka Metropolitan University, \\
Osaka 558-8585, Japan}
\\
${}^{b}${\it Faculty of Science and Engineering, Kyushu Sangyo University, \\
Fukuoka 813-8503, Japan}
\\
${}^{c}${\it Nambu Yoichiro Institute of Theoretical and Experimental Physics (NITEP), \\
Osaka Metropolitan University,
Osaka 558-8585, Japan}
\end{center}
\vspace*{1cm}

\begin{abstract}
We study a six-dimensional U(1)$_\chi$ gauge theory compactified on a magnetized torus, 
where the zero mode of the extra-dimensional gauge field (a Wilson-line (WL) scalar field) plays 
the role of a pseudo-Nambu-Goldstone (pNG) dark matter (DM) candidate. 
The pNG DM is naturally included by construction without introducing an additional scalar field. 
We show that the leading spin-independent DM-nucleus amplitude is suppressed by momentum transfer in our model 
as expected from the pNG DM model.
This suppression allows the model to evade the current severe direct-detection bounds 
while achieving the observed thermal relic abundance in well-defined regions of parameter space. 
\end{abstract}
\end{titlepage}

\section{Introduction}
Dark matter (DM) is supported by robust cosmological evidences, yet its particle identity remains unknown.
The Weakly Interacting Massive Particle (WIMP) framework is attractive because thermal freeze-out mechanism 
in the early universe fixes the relic abundance, and it has been long served as a standard benchmark.
Over the past decade, however, direct detection experiments have set 
very strong limits on the elastic scattering cross section with nucleons \cite{PandaX,XENONnT,LZ}.
Without a mechanism that naturally suppresses low energy DM-nucleon interactions while keeping thermal freeze-out, 
simple WIMP models become untenable.
Simply making the couplings very small tends to spoil consistency with the relic abundance, collider data, 
and cosmology.
Strategies that rely on accidental cancellations or strong fine tuning are not theoretically robust.

From this viewpoint, pseudo-Nambu-Goldstone (pNG) DM is a promising alternative \cite{GLT}.
An approximate shift symmetry protects the mass and the low energy interactions.
In the non-relativistic limit, the elastic scattering amplitude of the DM and the nucleon 
is suppressed as $\propto q^2$ with the momentum transfer $q$, 
thus naturally satisfying the latest direct detection limits ($t$-channel cancellation).
At the same time, annihilation at higher energy is not necessarily constrained by the shift symmetry, 
which helps to keep the value of $\braket{\sigma v}$ required by thermal freeze-out.
A conceptual issue remains concerning the origin of both the scalar field and the symmetry that gives rise to the NG boson.
Many pNG DM models have been developed and have substantially advanced this program;
see for example \cite{ATT,ORS,ATTY,ORST, Abe:2021vat, AH,LCJYZ,OSTUY,AHT,STT}.

Our approach is complementary to the origin of the scalar field and its protecting symmetry on a geometrical footing: flux compactifications.
Flux compactifications have a compact space like $T^2$ with non-trivial background magnetic flux.
Flux compactifications, developed in the context of string theory \cite{BKLS,IU}, 
give hints about the origin of generations \cite{Witten,ACKO} and the pattern of Yukawa couplings 
in the Standard Model (SM) \cite{CIM,ACKO2,HK,MS}.
In higher-dimensional theories with flux compactifications, the scalar field arises as the zero mode of a higher-dimensional gauge field (called Wilson-line (WL) scalar field), 
and the shift symmetry appears as the remnant of translations on the compact space \cite{Buchmuller:2016gib, BDD, Ghilencea:2017jmh, Honda:2019ema, Hirose:2019ywp, Hirose:2021rit, Akamatsu:2022jkg, KOT, Hirose:2024vvx, Kojima:2024yog, Franken:2022xmt}.
In the effective four-dimensional description, this symmetry is global, 
therefore the WL scalar field behaves as the NG boson of translations on the extra space, 
with no need to introduce an extra singlet scalar field or to postulate an ad hoc global symmetry.

As a candidate of the origin of pNG DM and its approximate shift symmetry, we consider a pNG DM model in six dimensional (6D) $\u 1_{\chi}$ gauge theories compactified on a two-dimensional torus with magnetic flux.
In our model, the WL scalar field plays the role of the pNG DM, and the translation of the extra space induces the shift symmetry.
We also search well-defined regions of parameter spaces consistent with the thermal relic abundance, unitarity bound, and the constraint of SM Higgs invisible decay.

The rest of this paper is organized as follows.
In section 2, we introduce a simple 6D model where the Higgs field is neutral under $\u1_\chi$ gauge symmetry 
and explains the $t$-channel cancellation mechanism. 
We show the parameter regions of this model, which satisfies the constraints for the relic abundance of the DM, 
the unitarity bound for the heavy Higgs field and the SM Higgs boson invisible decay width.
In section 3, we consider the model in which the Higgs field is charged under $\u1_\chi$ gauge symmetry.  
In this model, the interactions between the DM and Higgs field are present as a result of the $\u1_\chi$ gauge invariance.  
We also show the parameter regions of this model, which satisfies the constraint that the DM mass is smaller than the compactification scale in addition to the constraints discussed in the previous model. 
Section 4 presents our conclusions.

\section{A model of $\u1_\chi$ neutral Higgs boson}
As a preliminary step, we analyze a simple model and show that, 
in a 6D $\u1_\chi$ theory compactified on a torus with background magnetic flux, 
the WL scalar field exhibits the same $t$-channel cancellation as pNG DM \cite{GLT}, 
leading to momentum-suppressed scattering at small momentum transfer.

The 6D spacetime is $M^4 \times T^2$, 
where $M^4$ denotes Minkowski spacetime and $T^2$ is a two-dimensional square torus with an area $L^2$.
The 6D Lagrangian is
	\begin{align}
	\mc L_6&\supset-\frac{1}{4} F_{MN} F^{MN}-\frac{1}{2} \left(\partial_\mu A^\mu + \partial_m A^m\right)^2- |\partial_M H|^2-V(\phi,H)\notag \\
	&\supset-\frac{1}{4} F_{\mu\nu} F^{\mu\nu}
				-\partial_\mu\phi\partial^\mu\bar{\phi}
				- |\partial_M H|^2
				-V(\phi,H).
	\label{eq:lag}
	\end{align}
Here $M,N\in\{0,1,2,3,5,6\}$.
In the first line, the first term is the gauge kinetic term with the field strength tensor $F_{MN}=\partial_M A_N - \partial_N A_M$, 
and the second term is the gauge-fixing term where $\mu=0,1,2,3, m=5,6$.
The third term is the kinetic term of a complex scalar field $H$.
The last term constitutes the scalar potential, 
where the complex scalar field $\phi$ is defined 
as a linear combination of the extra-dimensional gauge field components $A_{5,6}$,
	\begin{align}
		\phi
		\equiv \frac{1}{\sqrt{2}} \left(A_6 + i A_5\right)
		\equiv \braket{\phi} + \varphi,
	\end{align}
where $\braket\phi$ is introduced by a magnetic flux which we will see later.
We refer to the zero mode of the fluctuation $\varphi$ as the WL scalar field.
The field $H$ is a complex scalar field, assumed to be neutral under $\u1_\chi$, 
and its zero mode is identified with the SM Higgs field.
We derive the kinetic term of the scalar field $\phi$ from the gauge kinetic term in the first line of Eq. \eqref{eq:lag}.

The potential $V(\phi,H)$ has
	\begin{align}
	V(\phi,H)=- \frac{m_H^2}{2} |H|^2
			+ L^2 \frac{\lambda_H}{2} |H|^4
			+ L^2 \lambda_{H\phi} |H|^2 |\phi|^2
			-\frac{m_\phi^2}{2} |\phi|^2
			+ L^2 \frac{\lambda_\phi}{2} |\phi|^4.
	\end{align}
The first two terms correspond to the usual Higgs potential.
The remaining three terms are inserted explicitly in this model and do not arise automatically from the gauge kinetic sector (we will later consider a setup where an interaction of the same form is generated through gauge couplings).
Among them, the $|H|^2|\phi|^2$ piece provides the Higgs portal, while the $|\phi|^2$ and $|\phi|^4$ pieces are included to induce a vacuum expectation value for $\phi$, thereby explicitly breaking torus translations and giving a mass to an imaginary part of the WL phase.
Although in many constructions a constant WL phase effectively plays the role of $\braket{\phi}$, here we model it by an explicit potential for simplicity.

We also introduce a background magnetic field (flux) in the extra-dimensional components of the gauge field,
\footnote{If we take $\braket{A_5} = - f x_6 + v_\chi / L$ and $\braket{A_6} = v_s / L$, 
the potential is minimized at $v_\chi = f L^2 / 2$, which is equivalent to $\braket{A_5} = - f ( x_6 - L/2 )$.}
	\begin{align}
		\braket{A_5} = - f \left(x_6 - \frac{L}{2}\right), \quad
		\braket{A_6} = \frac{v_s}{L}.
	\label{eq:flux}
	\end{align}
The flux is quantized (and related to the Landau-level degeneracy) as
	\begin{align}
		\frac{g}{2\pi} \int_{T^2}\dd^2 x~\braket{F_{56}}
		= \frac{g f L^2}{2\pi} = N \in \mathbb{Z},
	\label{eq:degeneracy}
	\end{align}
where $g$ is a 6D gauge coupling constant with mass dimension $-1$. 
For simplicity, we set $N=1$ hereafter.

In the absence of non derivative terms for $\phi$,  
the Lagrangian is invariant under torus translations $\delta_T = \epsilon_5 \partial_5 + \epsilon_6 \partial_6$, 
where $\epsilon_{5,6}$ are constant translation parameters on $T^2$. 
The effect of the flux is captured by the transformation of the fluctuation $\varphi$\cite{BDD},
	\begin{align}
		\delta_T \varphi
		= \left(\epsilon_5 \partial_5 + \epsilon_6 \partial_6\right)\varphi
			- \frac{i}{\sqrt{2}} f \epsilon_6 .
	\end{align}
Equivalently, the imaginary part of the WL scalar field, $\chi$, which is associated with the $x_5$ direction 
where the flux is introduced in Eq. \eqref{eq:flux}, shifts by a constant under translations on $T^2$.
It is therefore the NG boson of that symmetry, and we identify it as the DM candidate.
By introducing the potential terms for the scalar field $\phi$, 
the translational symmetry on the torus is explicitly broken,
and $\chi$ acquires a mass proportional to the symmetry breaking parameter.

To obtain the mass terms for the zero modes of the WL scalar field and the Higgs field,
the fields admit Kaluza-Klein (KK) expansions on $T^2$:
	\begin{align}
	\varphi(x_M)&=\sum_{l=-\infty}^\infty\sum_{m=-\infty}^\infty
		\varphi_{l,m}(x_\mu)\lambda_{l,m}(x_m),
	\label{eq:vpKK} \\
	H(x_M)&=\sum_{l=-\infty}^\infty\sum_{m=-\infty}^\infty
		H_{l,m}(x_\mu)\lambda_{l,m}(x_m),
	\end{align}
where the wave functions $\lambda_{l,m}$ are
	\begin{align}
	\lambda_{l,m}(x_m)=\frac1Le^{2i\pi(lx_5+mx_6)/L}.
	\end{align}
We redefine the zero modes as
	\begin{align}
	\varphi_0
	\equiv\frac1{\sqrt2}\left(s+i\chi\right),
		\quad
		H_0=\frac1{\sqrt2}(v_h+h)e^{i\pi/v_h}.
	\end{align}
After expanding the fields around the vacuum expectation values (VEVs) $v_s$, $f$ (see Eq. \eqref{eq:flux}) and $v_h$, and performing the integration on $T^2$ from 6D Lagrangian to 4D effective Lagrangian,
we obtain the quadratic terms and the cubic interactions relevant to the $t$-channel exchange amplitude:
	\begin{align}
	\int_{T^2}\dd^2x\left[\frac{m_\phi^2}2|\phi|^2\right]
		&\supset\frac{m_\phi^2}{4}\left[
			v_s^2+\frac1{12}f^2L^4
			+s^2+\chi^2
		\right],\\[3mm]
	\int_{T^2}\dd^2x\left[-L^2\frac{\lambda_\phi}2|\phi|^4\right]
		&\supset-\frac{\lambda_\phi}{8}\left[
			v_s^4+\frac16v_s^2f^2L^4
			+\frac1{80}f^4L^8+2\left(3v_s^2+\frac1{12}f^2L^4\right)s^2\right.
		\notag \\
		&\hspace{17mm}\left.
			+2\left(v_s^2+\frac14f^2L^4\right)\chi^2+4v_ss\chi^2
		\right],\\[3mm]
	\int_{T^2}\dd^2x\left[\frac{m^2_H}2|H|^2\right]
		&\supset\frac{m^2_H}4\left(v_h^2+h^2\right),\\[3mm]
	\int_{T^2}\dd^2x\left[-L^2\frac{\lambda_H}2|H|^4\right]
		&\supset-\frac{\lambda_H}8\left(v_h^4+6v_h^2h^2\right),\\[3mm]
	\int_{T^2}\dd^2x\left[-L^2\lambda_{H\phi}|H|^2|\phi|^2\right]
		&\supset-\frac{\lambda_{H\phi}}{4}\left[
			\left(v_s^2+\frac1{12}f^2L^4\right)\left(v_h^2+h^2\right)
			+4v_hv_shs
		\right.\notag \\
		&\hspace{20mm}\left.
			+v_h^2\left(s^2+\chi^2\right)
			+2v_hh\chi^2
		\right].
	\end{align}
	\begin{figure}[htpb]
	\centering
		\begin{tikzpicture}
		\begin{feynhand}
		\vertex[particle] (a) at (0,2) {$\chi$};
		\vertex[particle] (b) at (0,0) {$f$};
		\vertex (c) at (1.5,1.5);
		\vertex (d) at (1.5,0.5);
		\vertex[particle] (e) at (3,0) {$f$}; 
		\vertex[particle] (f) at (3,2) {$\chi$};
		\propag[sca] (a) to (c);
		\propag[sca] (f) to (c);
		\propag[sca] (c) to [edge label=$h_{1,2}$] (d) ;
		\propag[fer] (b) to (d);
		\propag[fer] (d) to (e);
		\end{feynhand}
		\end{tikzpicture}
		\caption{$t$-channel $\chi f \to \chi f$ scattering mediated by $h_{1,2}$.}
		\label{fig:t-channel}
	\end{figure}
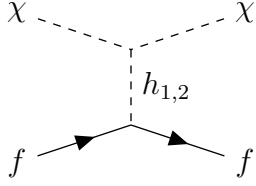
Note that the tree-level potential evaluated at the vacuum is given by
	\begin{align}
	\braket{V}
	&=-\frac{m_\phi^2}{4}\left(
			v_s^2+\frac1{12}f^2L^4
		\right)
		+\frac{\lambda_\phi}{8}\left[
			v_s^4+\frac1{6}v_s^2f^2L^4
			+\frac1{80}f^4L^8
		\right]\notag \\
	&\quad+\frac{\lambda_{H\phi}}{4}
		\left(v_s^2+\frac1{12}f^2L^4\right)v_h^2
		-\frac{m^2_H}4v_h^2
		+\frac{\lambda_H}8v_h^4,
	\end{align}
the minimization (stationary) conditions, $\partial\braket{V}/\partial v_{s,h}=0$ are
	\begin{align}
	m_\phi^2&=\lambda_\phi\left(v_s^2+\frac1{12}f^2L^4\right)
		+\lambda_{H\phi}v_h^2,\\
	m_H^2&=\lambda_Hv_h^2
		+\lambda_{H\phi}\left(v_s^2+\frac1{12}f^2L^4\right).
	\end{align}
Using these conditions, the quadratic mass terms for $h$, $s$, and $\chi$ can be written as
	\begin{align}
	\mc L&\supset
		-\frac12\left[\begin{array}{cc} h & s\end{array}\right]
		\left[\begin{array}{cc}
			\lambda_Hv_h^2 & \lambda_{H\phi}v_hv_s \\
			\lambda_{H\phi}v_hv_s & \lambda_\phi v_s^2
		\end{array}\right]
		\left[\begin{array}{c}
			h \\ s
		\end{array}\right]
		-\frac{1}{24}\lambda_\phi f^2L^4\chi^2\notag \\
	&\equiv-\frac12\left[\begin{array}{cc} h & s\end{array}\right]
		\mc M^2
		\left[\begin{array}{c}
			h \\ s
		\end{array}\right]
		-\frac{1}{24}\lambda_\phi f^2L^4\chi^2,
	\label{eq:mass_Q=0}
	\end{align}
from which $m_{\chi}^2=\lambda_\phi f^2L^4/12$ follows immediately.
The mass matrix $\mc M^2$ in the $[h,s]^t$ basis is diagonalized by the mixing angle
\begin{align}
	\tan2\theta=\frac{2\lambda_{H\phi}v_hv_s}{\lambda_\phi v_s^2-\lambda_Hv_h^2},
	\end{align}
which yields the eigenvalues
	\begin{align}
	m_{h_1,h_2}^2=\frac12\left\{
			\lambda_\phi v_s^2+\lambda_Hv_h^2
			\mp\sqrt{
				(\lambda_\phi v_s^2-\lambda_Hv_h^2)^2
				+4\lambda_{H\phi}^2v_h^2v_s^2
			}
		\right\}.
	\end{align}
$m_{h_1(h_2)}$ corresponds to the smaller (larger) mass eigenvalue 
and $h_{1,(2)}$ is identified with the SM (second) Higgs field, respectively.  
In the mass eigenstate basis, the cubic interactions relevant for the $t$-channel exchange (Fig. \ref{fig:t-channel}) are
	\begin{align}
	\mc L_{\chi\chi h_i}
		&=-\frac{\lambda_{H\phi}}2v_hh\chi^2
			-\frac{\lambda_\phi}2v_ss\chi^2\notag \\
		&=\frac1{2v_s}\chi^2
			\left(m_{h_1}^2\sin\theta h_1-m_{h_2}^2\cos\theta h_2\right).
	\end{align}
If the SM fermions ($f, \bar{f}$) carry no $\u1_\chi$ charge, their Yukawa interactions are identical to those in the SM:
	\begin{align}
	\mc L_{\mathrm{Yukawa}}&=-h\sum_f\frac{m_f}{v_h}\ol ff
	=-(\cos\theta h_1+\sin\theta h_2)\sum_f\frac{m_f}{v_h}\ol ff.
	\end{align}
The $t$-channel scattering amplitude is then
	\begin{align}
	\mc A_{t}&\propto
		-\frac{m_{h_1}^2\sin\theta\cos\theta}{t-m_{h_1}^2}
		+\frac{m_{h_2}^2\cos\theta\sin\theta}{t-m_{h_2}^2}
	=\sin\theta\cos\theta\frac{t(m_{h_2}^2-m_{h_1}^2)}{m_{h_1}^2m_{h_2}^2}
		+\mc O(t^2)
	\label{eq:amplitude}
	\end{align}
which cancels in the $t\to0$ limit as expected.

Next, we discuss the parameter space for the DM relic abundance to reproduce the DM density $\Omega_ch^2=0.12$\cite{Planck}.
We compute the thermally averaged annihilation cross sections $\braket{\sigma v}$ 
for $\chi\chi\to f\bar f,~W^+W^-,~ZZ,~h_i h_j~(i,j=1,2)$ and find the parameter regions 
realizing $\braket{\sigma v}\simeq3\times10^{-26}~\mathrm{cm^3/s}$.
In our analysis, we impose the constraints on the perturbative unitarity bound from high-energy $h_2 h_2\to h_2 h_2$ scattering, 
e.g. $\lambda_\phi<8\pi/3$, and the Higgs invisible width limit $\mathrm{Br}(h_1\to\mathrm{inv})\leq0.107$ \cite{ATLAS}.

\begin{figure}[htbp]
	\begin{minipage}[b]{0.5\linewidth}
	\centering
	\includegraphics[keepaspectratio, scale=0.8]{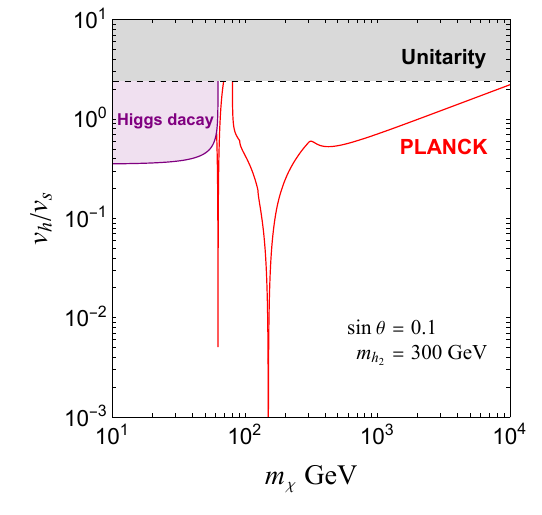}
	\subcaption{DM mass $m_\chi$ vs $v_h/v_s$}
	\end{minipage}
	\begin{minipage}[b]{0.5\linewidth}
	\centering
	\includegraphics[keepaspectratio, scale=0.8]{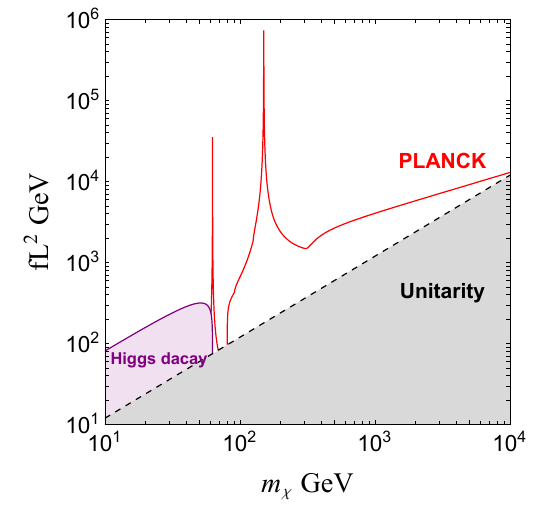}
	\subcaption{DM mass $m_\chi$ vs effective flux $fL^2$}
	\end{minipage}
	\caption{Allowed range of DM mass with $\sin\theta=0.1$ and $m_{h_2}=300$ GeV.}
	\label{fig:Q=0}
	\end{figure}

The results of our analysis are shown below (Fig. \ref{fig:Q=0}).
In the plots, parameter choices that reproduce the DM relic abundance are shown by a red curve;
the regions excluded by perturbative unitarity and the invisible Higgs decay are shaded gray and purple, respectively.
The model has four free parameters.
In the plots, we fix two of them, the mixing angle between $s$ and $h$, $\sin\theta$ 
and the second Higgs mass $m_{h_2}$, and scan the remaining two parameters to form a two-dimensional parameter plane.
Along the relic abundance curve, two resonances appear, 
corresponding to $s$-channel poles at $m_{\chi}=m_{h_1}/2=62.5~\mathrm{GeV}$ and $m_{\chi}=m_{h_2}/2$.

Since the parameter set of our model is identical to that of the pNG DM model in \cite{GLT}, 
the left panel of Fig. \ref{fig:Q=0} with vertical axis $v_h/v_s$ is the same as in the pNG DM case.
By contrast, the right panel of Fig. \ref{fig:Q=0} displays the same solutions with the 4D effective flux $fL^2$ on the vertical axis.
	\begin{align}
	fL^2=\frac{2\sqrt3v_hm_\chi}{\sqrt{\sin^2\theta m_{h_1}^2+\cos^2\theta m_{h_2}^2}}
		\left(\frac{v_h}{v_s}\right)^{-1}.
	\end{align}
From the flux quantization condition (Eq.~\eqref{eq:degeneracy}, $N=1$), one has $fL^2=2\pi/g$.
Hence regions with larger $fL^2$ correspond to smaller six-dimensional gauge coupling $g$.
Since the DM mass is given by the flux as $m_{\chi}^2=\lambda_{\phi} f^2 L^4/12$ (see Eq. \eqref{eq:mass_Q=0}), 
the two quantities $m_\chi$ and $fL^2$ are approximately linear away from the resonances, 
and their characteristic scales are comparable.
In the resonance regions $m_{\chi}\simeq m_{h_1}/2$ and $m_{\chi}\simeq m_{h_2}/2$, 
this proportionality is not valid. 
Accordingly, outside the resonance regions, the higher-dimensional viewpoint favors larger $m_{\chi}$, 
which enhances decoupling from KK modes and reduces loop sensitivity, 
whereas near the $s$-channel poles this correlation need not hold.
We found that the DM mass is allowed in a wide range between 60 GeV and 10 TeV.


\newpage
\section{A model of $\u1_\chi$ charged Higgs boson}
In the previous section, we introduced the portal interaction $|H|^2|\phi|^2$ by hand.
In a case that the SM field is charged under $\u1_\chi$, we can generate the interaction of the same type through gauge couplings.
	\begin{align}
	\mc L&\supset
		-\frac14F_{MN}F^{MN}
			-\frac12\left(\partial_\mu A^\mu+\partial_mA^m\right)^2
			+\frac{m_\phi^2}2|\phi|^2-L^2\frac{\lambda_\phi}2|\phi|^4
		\notag \\
	&\quad
		-|D_MH|^2+\frac{m^2_H}2|H|^2	
		-L^2\frac{\wt\lambda_H}2|H|^4.
	\end{align}
In this Lagrangian, the portal $|H|^2|\phi|^2$ is automatically contained in the Higgs kinetic term with the covariant derivative $D_M=\partial_M-igA_M$.
Adopting the same flux configuration as in the previous section, 
we define the creation and annihilation operators:
	\begin{align}
	a^\dagger
	&\equiv-\frac i{\sqrt{2gf}}\left[\partial_5-i\partial_6
		+igf\left(x_6-\frac L2\right)\right],\\[2mm]
	a&\equiv-\frac i{\sqrt{2gf}}\left[\partial_5+i\partial_6
		+igf\left(x_6-\frac L2\right)\right]. 
	\end{align}
We then introduce the Landau-level wavefunctions $\xi_{n,j}$ \cite{CIM,HK,KOT} satisfying
	\begin{align}
	\left[a,a^\dagger\right]=1,\quad
	a\xi_{0,j}=0,
	\end{align}
together with the ladder relations
	\begin{align}
	a\xi_{n,j}=\sqrt n\xi_{n-1,j},\quad
	a^\dagger\xi_{n,j}=\sqrt{n+1}\xi_{n+1,j},
	\end{align}
and the orthonormality condition
	\begin{align}
	\int_{T^2}\dd^2x~\ol\xi_{m,j}\xi_{n,k}=\delta_{m,n}\delta_{j,k}.
	\end{align}
Using these wavefunctions, the Higgs field admits the KK expansion
	\begin{align}
	H(x_M)=\sum_{n=0}^\infty\sum_{j=0}^{N-1}
		H_{n,j}(x_\mu)\xi_{n,j}(x_m),\quad
	(\mu=0,1,2,3,~m=5,6,)
	\end{align}
where $n$ labels the Landau level and $j$ denotes the ground-state degeneracy.
Among the KK modes, we regard the zero mode $(n=0)$ as the SM Higgs field and expand it as
	\begin{align}
	H_{0,0}(x_\mu)=\frac1{\sqrt2}(v_h+h(x_\mu))e^{i\pi(x_\mu)/v}.
	\end{align}
Note that, since $\u1_\chi$ is Abelian and $\vp$ does not interact with the background flux, 
the KK expansion of $\vp$ is same as Eq. \eqref{eq:vpKK}; we define the zero mode as $\vp_0=(s+i\chi)/\sqrt2$.

We now perform the integration on $T^2$ from 6D Lagrangian to 4D Lagrangian, and extract the mass and interaction terms for the zero modes.
The kinetic term is calculated as
	\begin{align}
	&\int_{T^2}\dd^2x\left[-|D_mH|^2\right]\notag \\
	&=-\frac12\int_{T^2}\dd^2x\left[
		\left|\sum_{n,j}\left\{i\sqrt{2gf}\sqrt{n}H_{n-1,j}
			-g\left(\sqrt2\vp+\frac{v_s}L\right)H_{n,j}\right\}\xi_{n,j}\right|^2\right.\notag \\
	&\hspace{30mm}\left.+\left|\sum_{n,j}\left\{-i\sqrt{2gf}\sqrt{n+1}H_{n+1,j}
		-g\left(\sqrt2\ol\vp+\frac{v_s}L\right) H_{n,j}\right\}\xi_{n,j}\right|^2\right].
	\end{align}
Extracting only the zero mode $\vp_0$, one finds
	\begin{align}
	&\int_{T^2}\dd^2x\left[-|D_mH|^2\right]\notag \\
	&\supset-\frac12\sum_{n,j}\left[
		\left|i\sqrt{2gf}\sqrt{n}H_{n-1,j}-\frac gL\left(\sqrt2\vp_0+v_s\right)H_{n,j}\right|^2
		\right.\notag \\
	&\hspace{23mm}\left.+\left|-i\sqrt{2gf}\sqrt{n+1}H_{n+1,j}-\frac gL\left(\sqrt2\ol\vp_0+v_s\right)H_{n,j}\right|^2\right]
		\notag \\
	&=-\sum_{n,j}\left[2gf\left(n+\frac12\right)|H_{n,j}|^2
		+\left(\frac gL\right)^2|\sqrt2\vp_0+v_s|^2|H_{n,j}|^2\right.\notag \\
	&\hspace{20mm}\left.+i\frac{g\sqrt{2gf(n+1)}}L\left\{\left(\sqrt2\vp_0+v_s\right)H_{n+1,j}\ol H_{n,j}
			-c.c.\right\}
		\right].
	\end{align}
The first term represents the KK masses of $H_{n,j}$ generated by the flux.
The second term includes precisely the quartic interaction that couples the dark sector to the Standard Model.
The third term encodes a cubic interaction between the DM and the Higgs field; due to the background flux, 
the selection rules restrict it to couplings between adjacent KK modes.
Due to the cubic interaction, this model may have a possibility that the decay channel $\varphi_0\to H_{1,0}H_{0,0}$ is kinematically open if the compactification scale is around TeV scale.
Taking this decay channel into account, the stable DM must be regarded as the first Higgs KK mode $H_{1,0}$, not the WL scalar $\varphi_0$.
Since we consider the DM as $\varphi_0$, we forbid this decay and require the DM to be lighter than the first excited Higgs mode, $m_{\chi}\lesssim m_{\mathrm{KK}}$.
Including the Higgs potential, the mass of $H_{1,0}$ is approximately set by the Landau gap, 
and we take $m_{\mathrm{KK}}^2\sim 2gf$; accordingly, we impose the conservative bound
	\begin{align}
	m_\chi\lesssim\sqrt{2gf}
	=\frac{2\sqrt\pi}L,\quad
	(\because \mathrm{Eq.}~\eqref{eq:degeneracy}.) 
	\label{eq:bound}
	\end{align}
After extracting the zero mode, we obtain
	\begin{align}
	&\int_{T^2}\dd^2x\left[-|D_mH|^2\right]\notag \\
	&\supset-2gf\left(0+\frac12\right)|H_{0,0}|^2
		-\left(\frac gL\right)^2|s+i\chi+v_s|^2|H_{0,0}|^2\notag \\
	&=-\frac12gf\left(v_h^2+h^2\right)
		-\frac12\left(\frac gL\right)^2v_s^2\left(v_h^2+h^2\right)\notag \\
	&\quad-\frac12\left(\frac gL\right)^2\left(s^2+\chi^2\right)v_h^2
		-2\left(\frac gL\right)^2v_hv_shs
		-\left(\frac gL\right)^2v_h\chi^2h.
	\end{align}
Next, we compute the remaining terms.
The $|\phi|^2$ and $|\phi|^4$ terms are the same as in the previous section.
The $|H|^2$ term follows directly from the orthonormality of the Landau-level wavefunctions:
	\begin{align}
	\int_{T^2}\dd^2x~\frac{m^2_H}2|H|^2
	&=\frac{m^2_H}2\sum_{n,j}|H_{n,j}|^2
	\supset
		\frac{m^2_H}4\left(v_h^2+h^2\right)
	\end{align}
For the $|H|^4$ term, keeping only the pieces built from the zero mode $H_{0,0}$, we find
	\begin{align}
	&\int_{T^2}\dd^2x\left[-L^2\frac{\wt\lambda_H}2|H|^4\right]
	\supset-L^2\left(\int_{T^2}\dd^2x~|\xi_{0,0}|^4\right)\frac{\wt\lambda_H}2|H_{0,0}|^4.
	\end{align}
The torus integral evaluates to
	\begin{align}
	\int_{T^2}\dd^2x~|\xi_{0,0}|^4
		=\frac{\sqrt N}{L^2}\left(\vartheta_3\left(0,e^{-\pi N}\right)\right)^2
		\equiv\frac{\theta_N}{L^2},
		\quad\left(\vartheta_3(z,q)=\sum_{n\in\mathbb Z}q^{n^2}e^{2inz},\right)
	\end{align}
where $\vartheta_3(z,q)$ is the third Jacobi theta function (see Appendix \ref{app:xi4}).
Therefore, the vacuum and quadratic terms from the quartic interaction are
	\begin{align}
	&\int_{T^2}\dd^2x\left[-L^2\frac{\wt\lambda_H}2|H|^4\right]
	\supset-\theta_N\wt\lambda_H\left(\frac18v_h^4+\frac34v_h^2h^2\right)
	\end{align}
and we redefine the coupling constant as
	\begin{align}
	\theta_N\wt\lambda_H\equiv\lambda_H.
	\end{align}

Collecting the vacuum energy contribution, the mass terms and the cubic interactions, we obtain
	\begin{align}
	\mc L
	&\supset\frac{m_\phi^2}{4}\left(
			v_s^2+\frac1{12}f^2L^4
			\right)
		-\frac{\lambda_\phi}{8}\left(
			v_s^4+\frac{v_s^2}{6}f^2L^4
			+\frac1{80}f^4L^8
			\right)\notag \\
	&\quad-\frac12gfv_h^2-\frac12\left(\frac gL\right)^2v_s^2v_h^2
		+\frac{m_H^2}4v_h^2-\frac{\lambda_H}8v_h^4\notag \\
	&\quad+\frac14\left[
		m_\phi^2-\lambda_\phi\left(3v_s^2+\frac1{12}f^2L^4\right)
		-2\left(\frac gL\right)^2v_h^2
		\right]s^2\notag \\
	&\quad+\frac14\left[
		m_\phi^2-\lambda_\phi\left(v_s^2+\frac1{4}f^2L^4\right)
		-2\left(\frac gL\right)^2v_h^2
		\right]\chi^2\notag \\
	&\quad+\frac14\left[
		m_H^2-2gf-3\lambda_Hv_h^2
		-2\left(\frac gL\right)^2v_s^2
		\right]h^2
		-2\left(\frac gL\right)^2v_hv_shs\notag \\
	&\quad-\frac12\left[
		\lambda_\phi v_ss+2\left(\frac gL\right)^2v_hh
		\right]\chi^2.
	\end{align}
Solving the stationary conditions of the vacuum energy
	\begin{align}
	\braket{V}&=-\frac{m_\phi^2}{4}\left(v_s^2+\frac1{12}f^2L^4\right)
		+\frac{\lambda_\phi}{8}\left(v_s^4+\frac{v_s^2}{6}f^2L^4
		+\frac1{80}f^4L^8\right)\notag \\
	&\quad+\frac12gfv_h^2+\frac12\left(\frac gL\right)^2v_s^2v_h^2
		-\frac{m_H^2}4v_h^2+\frac{\lambda_H}8v_h^4
	\end{align}
with respect to $v_s$ and $v_h$, we obtain
	\begin{align}
	m_\phi^2&=\lambda_\phi\left(v_s^2+\frac1{12}f^2L^4\right)+2\left(\frac gL\right)^2v_h^2,\\
	m_H^2&=2gf+\lambda_Hv_h^2+2\left(\frac gL\right)^2v_s^2.
	\end{align}
Using these conditions, the quadratic part of the potential can be written as
	\begin{align}
	V&\supset\frac12\left[\begin{array}{cc} h & s\end{array}\right]
		\left[\begin{array}{cc}
			\lambda_Hv_h^2 & 2(g/L)^2v_hv_s \\
			2(g/L)^2v_hv_s & \lambda_\phi v_s^2
		\end{array}\right]
		\left[\begin{array}{c}
			h \\ s
		\end{array}\right]+\frac1{24}\lambda_\phi f^2L^4\chi^2.
	\end{align}
The mass matrix $\mc M^2$ in the $[h,s]$ basis is diagonalized by the mixing angle	
\begin{align}
	\tan2\theta=\frac{4(g/L)^2v_hv_s}{\lambda_\phi v_s^2-\lambda_Hv_h^2},
	\end{align}
which yields the eigenvalues
	\begin{align}
	m_{h_1,h_2}^2=\frac12\left\{\lambda_\phi v_s^2+\lambda_Hv_h^2
		\mp\sqrt{(\lambda_\phi v_s^2-\lambda_Hv_h^2)^2
			+16\left(\frac gL\right)^4v_h^2v_s^2}\right\}.
	\end{align}
In the mass eigenstate basis, the cubic interactions relevant for the $t$-channel exchange take the form
	\begin{align}
	\mc L_{\chi\chi h_i}
	&=-\left(\frac{g}L\right)^2v_hh\chi^2
		-\frac{\lambda_\phi}2v_ss\chi^2\notag \\
	&=\frac1{2v_s}\chi^2
		\left(m_{h_1}^2\sin\theta h_1-m_{h_2}^2\cos\theta h_2\right).
	\end{align}

We now turn to the fermion sector.
If SM fermions have Yukawa couplings to the Higgs, 
$\u1_\chi$ gauge invariance of the Yukawa terms requires the assignments of suitable $\u1_\chi$ charges to the fermions.
The fermion kinetic terms then have covariant derivatives, 
inducing an effective WL scalar-fermion cubic interaction through the extra-dimensional gauge components \cite{BDD}.
In a flux compactification, however, this vertex is off diagonal in the Landau-level index, 
therefore the zero-mode to zero-mode coupling vanishes.
Consequently, for zero modes the effect is suppressed by KK mixing and the KK mass scale, 
and a direct WL-fermion-fermion coupling can be neglected.
From the viewpoint of two-body decays, 
imposing the bound $m_{\chi}\lesssim m_{\mathrm{KK}}$ in Eq.\eqref{eq:bound} keeps such channels kinematically closed.
The $t$-channel exchange relevant for direct detection is therefore dominated by the Standard Model Yukawa $h$-$f$-$f$ interaction, 
preserving the momentum-transfer $q^2$ suppression of the amplitude (as in Eq. \eqref{eq:amplitude}).

\begin{figure}[htbp]
	\begin{minipage}[t]{0.5\linewidth}
	\centering
	\includegraphics[keepaspectratio, scale=0.8]{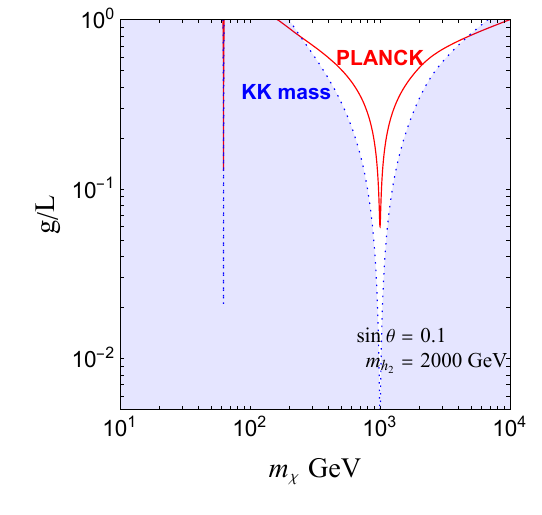}
	\subcaption{DM mass vs effective gauge coupling $g/L$}
	\end{minipage}
	\begin{minipage}[t]{0.5\linewidth}
	\centering
	\includegraphics[keepaspectratio, scale=0.77]{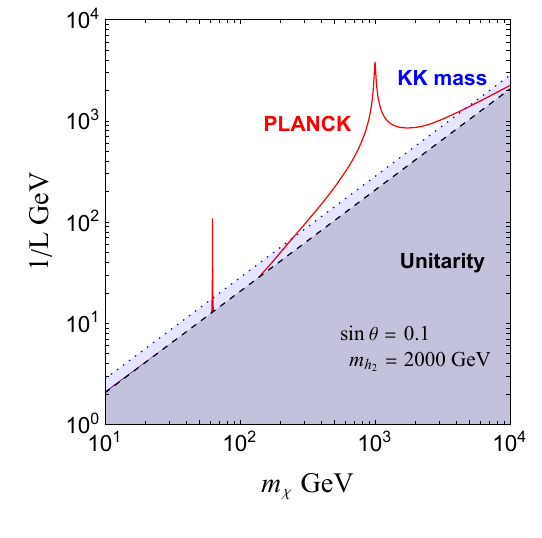}
	\subcaption{DM mass vs compactification scale $1/L$}
	\end{minipage}
	\caption{Allowed range of DM mass with $\sin\theta=0.1$ and $m_{h_2}=2000$ GeV.}
	\label{fig:Q>0_2TeV}
	\end{figure}

Our results in this section are given.
In the charged model, the portal coupling involves the gauge coupling $g$ and the DM mass depends on the effective flux $fL^{2}$;
together with the flux quantization condition \eqref{eq:degeneracy}, 
this allows us to determine the compactification scale $1/L$ explicitly.
	\begin{align}
	\frac{g}{L}&=\sqrt{\frac{\sin2\theta(m_{h_2}^2-m_{h_1}^2)}{4v_h^2}\cdot\frac{v_h}{v_s}},\\[2mm]
	\frac1L&=\frac{\sqrt6}{2\pi}\sqrt{\frac{\tan\theta(m_{h_2}^2-m_{h_1}^2)}{\tan^2\theta m_{h_1}^2+m_{h_2}^2}}
	m_\chi\left(\frac{v_h}{v_s}\right)^{-1/2}.
	\end{align}
We have to take into account a KK bound \eqref{eq:bound} 
in addition to the perturbative unitarity and Higgs invisible decay constraints.
Specifically, we require
	\begin{align}
	\frac1L>\frac{m_\chi}{2\sqrt{\pi}}.
	\end{align}
	\begin{figure}[htbp]
	\begin{minipage}[t]{0.325\linewidth}
	\centering
	\includegraphics[keepaspectratio, scale=0.6]{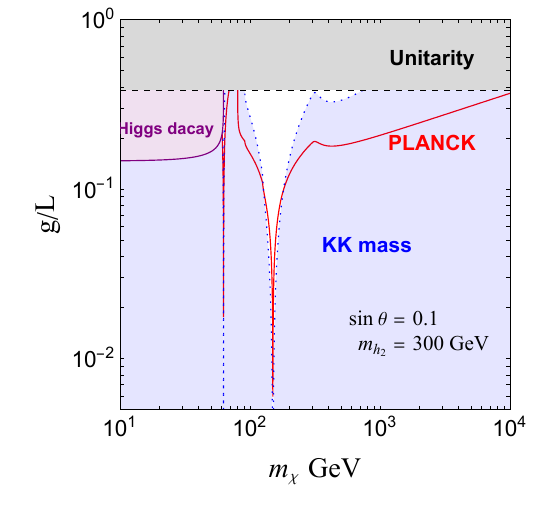}
	\subcaption{DM mass vs effective gauge coupling $g/L$}
	\label{fig:Q>0_300GeV_a}
	\end{minipage}
	\begin{minipage}[t]{0.325\linewidth}
	\centering
	\includegraphics[keepaspectratio, scale=0.6]{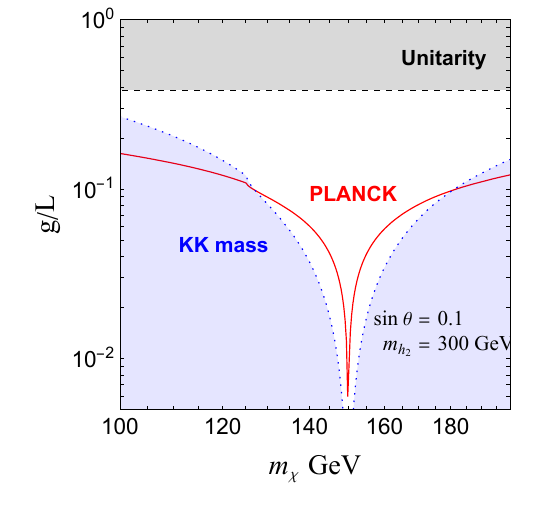}
	\subcaption{Zoom-in of Fig. \ref{fig:Q>0_300GeV_a}}
	\end{minipage}
	\begin{minipage}[t]{0.325\linewidth}
	\centering
	\includegraphics[keepaspectratio, scale=0.5675]{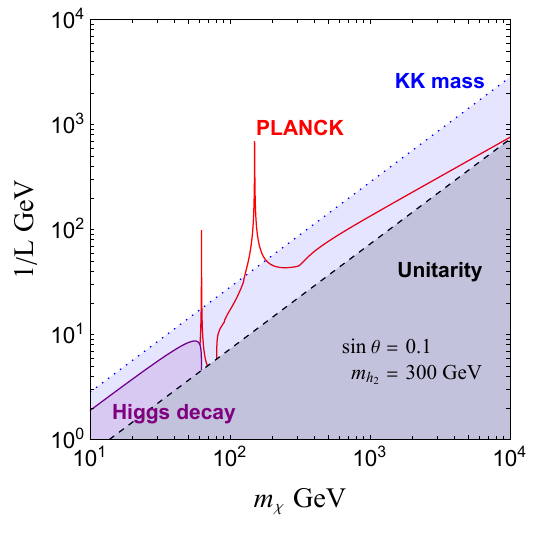}
	\subcaption{DM mass vs compactification scale $1/L$}
	\end{minipage}
	\caption{Allowed range of DM mass with $\sin\theta=0.1$ and $m_{h_2}=300$ GeV.}
	\label{fig:Q>0_300GeV}
	\end{figure}
In Fig. \ref{fig:Q>0_300GeV}, the parameter planes are displayed for $\sin\theta=0.1$ and $m_{h_2}=300~\mathrm{GeV}$ 
in terms of the flux compactification quantities.
The added blue region is excluded by the KK mass bound, which significantly reduces the range of viable DM masses ($130~\mathrm{GeV}<m_\chi<180~\mathrm{GeV}$).
The vertical axis of the left panel represents the effective four-dimensional $\u1_\chi$ gauge coupling $g/L$, and the central panel shows the zoom-in of it.
We find $g/L\ll 1$ throughout the viable region, therefore perturbation theory remains reliable.
In the right panel, shows, 
the compactification scale $1/L$ is taken as the vertical axis instead.
We see that the allowed compactification scale $1/L$ is a few hundred GeV 
for the choice of $\sin \theta=0.1, m_{h_2} =300~\mathrm{GeV}$, which is clearly lower than the experimental bound. 
On the other hand, increasing $m_{h_2}$ to the TeV scale (see Fig.~\ref{fig:Q>0_2TeV} with $m_{h_2}=2000~\mathrm{GeV}$) 
requires a larger $v_s$ to reproduce the relic abundance, and the contours shift upward.
Since the KK mass bound does not depend on $m_{h_2}$, the compactification scale $1/L$ consequently increases, 
the experimental bound for the compactification scale can be satisfied. 
We find that the DM mass is again allowed in a wide range between 250 GeV and 4700 GeV, a characteristic feature of pNG DM.

\section{Summary}
We have shown that, in a six-dimensional $\u1_\chi$ gauge theory with a background magnetic flux, 
the WL scalar field can be regarded as pseudo-Nambu-Goldstone dark matter (pNG DM) 
once the translational symmetry of the extra space is explicitly broken.

First, we analyzed a model in which the SM Higgs field is neutral under $\u1_\chi$ gauge symmetry 
and the interaction between the DM and the Higgs field is introduced by hand. 
We then computed the DM-fermion scattering amplitude 
and demonstrated that the leading contribution is proportional to the square of the momentum transfer ($t$-channel cancellation), 
confirming that the WL scalar field can behave as the pNG DM.
We also analyzed the parameter regions reproducing the observed relic abundance of DM 
and current constraints on the perturbative unitarity bound from high-energy $h_2 h_2\to h_2 h_2$ scattering, 
e.g. $\lambda_\phi<8\pi/3$, and the Higgs invisible width limit $\mathrm{Br}(h_1\to\mathrm{inv})\leq0.107$.
We found that the DM mass is allowed in a wide range between 60 GeV and 10 TeV. 

Next, we considered a model in which the Higgs field carries a charge under the $\u1_\chi$ gauge symmetry.
In this case, the portal potential does not need to be introduced by hand: 
the interaction arises naturally from the Higgs kinetic term.
As in the previous model, 
the DM-fermion $t$-channel amplitude cancels at small momentum transfer, 
therefore the WL scalar field can be similarly identified as the pNG DM.
We presented parameter regions consistent with the relic abundance and experimental limits.
In this model, higher-dimensional information enters in two places: the DM mass and the portal coupling.
These two quantities allow us to infer the gauge coupling and the compactification scale.
We find that the effective four-dimensional gauge coupling is generally less than one, 
which means that perturbation theory remains valid.
For a commonly used benchmark the second Higgs mass $m_{h_2}$ of 300 GeV, 
the DM mass is allowed in a narrow range between 130 GeV and 180 GeV, and
the compactification scale is around a few hundred GeV, which is excluded from the collider experimental bound.
However,  if we increase $m_{h_2}$ to be TeV scale, 
we find that the DM mass is again allowed in a wide range between 250 GeV and 4700 GeV, and
the resulting compactification scale can be found around TeV scale accordingly and satisfy the experimental bound.

Finally, it would be interesting to take into account the nonzero KK mode contributions in our analysis 
although we focused on the leading effects of zero mode contributions in this paper. 
Our prediction of the cross section for the DM direct detection would be obtained from such study. 
This issue is left for our future work.

\section*{Acknowledgments}
T.H. would like to thank Yoshiki Uchida for Mathematica code prototype of Fig. \ref{fig:Q=0},\ref{fig:Q>0_300GeV}, 
and Riasat Sheikh for valuable discussions.
This work was supported by JST SPRING, Grant Number JPMJSP2139 (K.A. and A.N.)
and
Japan Society for the Promotion of Science, Grants-inAid for Scientific Research, No. 25K07304 (N.M.)
and KSU Fundamental Research Fund in Kyushu Sangyo University, K024023 (T.H.).

\appendix
\section{Wave function}
\label{app:xi4}
According to \cite{KOT}, the wavefunctions $\xi_{n,j}$ in the Landau gauge are known to be 
	\begin{align}
	\xi_{n,j}
	&=\frac{\mc N_n}L\sum_{a\in\mathbb Z}e^{2i\pi N(x_5/L)(j/N+a)\}}
		e^{-\pi N\{(x_6-L/2)/L+j/N+a\}^2}\notag \\
	&\hphantom{=\frac{\mc N_n}L\sum_{a\in\mathbb Z}}
		\times H_n\left(\sqrt{2\pi N}
		\left(\frac{x_6-L/2}L+\frac{j}{N}+a\right)\right),
	\end{align}
where the dimensionless normalization factor $\mc N_n$ is given by
	\begin{align}
	\mc N_{n}=\frac{(2N)^{1/4}}{(2^nn!)^{1/2}}. 
	\end{align}

In the following, we evaluate the nontrivial wavefunction integrals on $T^2$.
	\begin{align}
	\int_{T^2}\dd^2x~|\xi_{0,0}|^4
	&=\int_0^L\dd x_5\int_0^L\dd x_6\left|\frac{\mc N_0}L\sum_{a\in\mathbb Z}
		e^{2i\pi N(x_5/L)a}e^{-\pi N(x_6/L-1/2+a)^2}\right|^4\notag \\
	&=\int_0^L\dd x_5\int_0^L\dd x_6~\frac{\mc N_0}L\sum_{a\in\mathbb Z}
		e^{2i\pi N(x_5/L)a}e^{-\pi N(x_6/L-1/2+a)^2}\notag \\
	&\hspace{30mm}\times\frac{\mc N_0}L\sum_{b\in\mathbb Z}
		e^{-2i\pi N(x_5/L)b}e^{-\pi N(x_6/L-1/2+b)^2}\notag \\
	&\hspace{30mm}\times\frac{\mc N_0}L\sum_{c\in\mathbb Z}
		e^{2i\pi N(x_5/L)c}e^{-\pi N(x_6/L-1/2+c)^2}\notag \\
	&\hspace{30mm}\times\frac{\mc N_0}L\sum_{d\in\mathbb Z}
		e^{-2i\pi N(x_5/L)d}e^{-\pi N(x_6/L-1/2+d)^2}.
	\end{align}
After performing the $x_5$ integration, we obtain
	\begin{align}
	\int_0^L\dd x_5~e^{2i\pi N(x_5/L)(a-b+c-d)}
	=L\delta_{a+c,b+d}.
	\end{align}
Then,
	\begin{align}
	\int_{T^2}\dd^2x~|\xi_{0,0}|^4
	&=\frac{2N}{L^2}\int_0^1\dd Y\sum_{a,b,c,d\in\mathbb Z}\delta_{a+c,b+d}~
		e^{-\pi N\left(Y-1/2+a\right)^2-\pi N\left(Y-1/2+b\right)^2}\notag \\
	&\hphantom{=\frac{2N}{L^2}\int_0^1\dd Y\sum_{a,b,c,d\in\mathbb Z}~}
		\times e^{-\pi N\left(Y-1/2+c\right)^2-\pi N\left(Y-1/2+d\right)^2}\notag \\
	&\hspace{6cm}\left(\mc N_n=\frac{(2N)^{1/4}}{(2^nn!)^{1/2}},\quad Y=\frac{x_6}L\right)\notag \\
	&=\frac{2N}{L^2}\sum_{a,b,c\in\mathbb Z}\int_0^1\dd Y
		e^{-\pi N\left(Y-1/2+a\right)^2-\pi N\left(Y-1/2+b\right)^2-\pi N\left(Y-1/2+c\right)^2-\pi N\left(Y-1/2+a-b+c\right)^2}\notag \\
	&=\frac{2N}{L^2}\sum_{a',b'\in\mathbb Z}\int_{\mathbb R}\dd Y'
		e^{-\pi N\left(Y'+a'\right)^2-\pi N\left(Y'+b'\right)^2-\pi NY'^2-\pi N\left(Y'+a'-b'\right)^2}\notag \\
	&\hspace{5cm}\left(Y-\frac12+c=Y',\quad a=a'+c,\quad b=b'+c\right)\notag \\
	&=\frac{2N}{L^2}\sum_{a',b'\in\mathbb Z}\int_{\mathbb R}\dd Y'
		e^{-4\pi N(Y'+a'/2)^2-\pi N\{b'^2+(a'-b')^2\}}\notag \\
	&=\frac{\sqrt N}{L^2}\sum_{s\in\mathbb Z}e^{-\pi Ns^2}
		\sum_{t\in\mathbb Z}e^{-\pi Nt^2}\quad
		(s\equiv b',~t\equiv a'-b',~\mathrm{so}~(s,t)\in\mathbb Z^2)
		\notag \\
	&=\frac{\sqrt N}{L^2}\left(\sum_{s\in\mathbb Z}e^{-\pi Ns^2}\right)^2
		=\frac{\sqrt N}{L^2}\left(\vartheta_3\left(0,e^{-\pi N}\right)\right)^2.
	\end{align}
where $\vartheta_3(z,q)$ is the third Jacobi theta function.

\let\doi\relax
\bibliographystyle{utphys28mod}
\bibliography{Ref}

\end{document}